\documentstyle[12pt]{article}

\newcommand{\quark}{Q_\alpha^A}
\newcommand{\antiquark}{\bar{Q}^\alpha_A}
\newcommand{\quarksix}{Q_\alpha^6}
\newcommand{\antiquarksix}{\bar{Q}^\alpha_6}
\newcommand{\higgs}{H_A}
\newcommand{\antihiggs}{\bar{H}^A}

\newcommand{\sigmaab}{\Sigma^A{}_{\! B}}
\newcommand{\Meson}{M^A{}_{\! B}}
\newcommand{\meson}{M^a{}_{\! b}}
\newcommand{\baryon}{B_a}
\newcommand{\antibaryon}{\bar{B}^b}

\newcommand{\su}[2]{SU(#1)_{#2}}
\newcommand{\sugut}{\su{5}{GUT}}
\newcommand{\uone}[1]{U(1)_{#1}}

\newcommand{\rep}[1]{{\bf #1}}

\newfont{\bg}{cmr10 scaled\magstep4}

\newcommand{\bigzerou}{\smash{\lower1.7ex\hbox{\bg 0}}}

\begin{document}
\title{
\vspace*{-4ex} \hfill {\large UT-717 \ \ \ } \\
  \vspace{5ex}
  Dynamical Models for Light Higgs Doublets \\
  in Supersymmetric Grand Unified Theories}
\author{T. Hotta, Izawa K.-I. and T. Yanagida\\
  \\  Department of Physics, University of Tokyo \\
  Bunkyo-ku, Tokyo 113, Japan}
\date{August, 1995}
\maketitle

\setlength{\baselineskip}{3.6ex}
\begin{abstract}
  We consider a supersymmetric hypercolor gauge theory with six flavors
  of quarks interacting strongly at the grand unification scale.
  Dynamical breaking of the grand unified $\sugut$ produces a massless
  pair of composite color-triplet states. A use of the missing partner
  mechanism yields eventually a pair of massless Higgs doublets,
  giving large masses to the color-triplet partners.
  We prove that this pair of massless Higgs doublets
  survives quantum corrections and remains in the low-energy spectrum as
  far as the supersymmetry is unbroken.
  Hence this solves the
  most serious problem -- doublet-triplet splitting in the grand
  unified theories.
  We also show that the
  dangerous dimension-five operators for nucleon decays are suppressed
  simultaneously, which makes our model yet more attractive.
\end{abstract}

\newpage
\section{Introduction}

The grand unified theory (GUT) has attracted us for a long time since
it was constructed in 1974 \cite{1}, because of its various interesting
and promising features.
In particular, the recent high-precision measurements on the Weinberg
angle have shown a remarkable agreement \cite{2} with the prediction of a
supersymmetric (SUSY) extension \cite{3} of the GUT.
The success of the SUSY-GUT has, thus, led us to a serious
reconsideration of SUSY-GUT models.

In the minimum SUSY-GUT, one must require an extremely precise
adjustment of parameters in order to obtain a pair of light Higgs
doublets \cite{3}. This pair of light Higgses is an inevitable
ingredient in the SUSY standard model. Therefore, the necessity of the
fine-tuning of parameters seems to be a crucial drawback in the
minimum SUSY-GUT.

There have been, in fact, several attempts \cite{4, 5, 6, 7, 8} to
have the light Higgs doublets without requiring the fine-tuning of
parameters.
In Ref.\cite{8} it has been shown that the dynamics of SUSY QCD-like
theory at the GUT scale naturally generates pairs of Higgs doublets at
low energies. However, quantum effects have not been fully
investigated in Ref.\cite{8} and seven flavors of (hyper)quarks have been
assumed in order to have a manifest symmetry to protect massless Higgs
doublets from quantum corrections. Although this model is very
interesting due to its natural accommodation of the Peccei-Quinn symmetry,
the low-energy spectrum is rather involved with two
pairs of light Higgs doublets.

In this paper we investigate the quantum dynamics of this QCD-like theory
and show that a pair of massless Higgs doublets in the
minimal model with six (hyper)quarks
is indeed stable quantum mechanically. This makes the dynamical approach
proposed in Ref.\cite{8} yet more attractive. We also stress that the
dangerous dimension-five operators for the nucleon decays are suppressed
simultaneously in the present model.

\section{Classical vacua and massless states}

Our model is based on a supersymmetric hypercolor $\su{3}{H} \times
\uone{H}$ gauge theory \cite{8} with $N_f$ flavors of quarks
$\quark$ in the $\rep{3}$ representation and antiquarks
$\antiquark$ in the $\rep{3}^\ast$ representation of $\su{3}{H}$, where
$\alpha = 1, \cdots, 3$ and $A = 1, \cdots, N_f$.
The chiral multiplets $\quark$ and $\antiquark$ have $\uone{H}$ charges $+1$
and $-1$, respectively.

Before investigating the realistic case of $N_f = 6$,
we first consider the basic case of $N_f = 5$. The anomaly-free
flavor symmetry is then given by
\begin{equation}
  \label{1}
  \su{5}{L} \times \su{5}{R} \times \uone{V} \times \uone{R} ,
\end{equation}
under which the quark multiplets transform as
\begin{equation}
  \label{2}
  \begin{array}{rl}
    \displaystyle
    \quark : & ( \rep{5}, \rep{1}, +1, \frac{2}{5}) \\
    \noalign{\vskip 1ex}
    \antiquark : & ( \rep{1}, \rep{5}, -1, \frac{2}{5}) \\
    \noalign{\vskip 1ex}
    & (\alpha = 1, \cdots, 3; A = 1, \cdots, 5) \; .
  \end{array}
\end{equation}
We note that $\uone{V}$ is nothing other than $\uone{H}$,
but we regard it as a flavor group.
The GUT gauge group $\sugut$ is also embedded in a part of the flavor
group $\su{5}{L} \times \su{5}{R}$. Namely, the quarks $\quark$ and
$\antiquark$ transform as $\rep{5}^\ast$ and $\rep{5}$ under
$\sugut$, respectively.

When $\sugut$ is spontaneously broken down to the standard gauge
group, there appear unwanted Nambu-Goldstone multiplets. To avoid
them, we introduce a chiral multiplet $\sigmaab$ in the adjoint
$(\rep{24})$ representation of $\sugut$ \cite{8}. Then, we have a
superpotential
\begin{equation}
  \label{3}
  W = \antiquark ( m \, \delta^A{}_{\! B} + \lambda \sigmaab) Q_\alpha^B +
  \frac{1}{2} \, m_\Sigma \, Tr(\Sigma^2),
\end{equation}
where $m$ and $m_\Sigma$ denote mass parameters.
Here, we have omitted a $Tr(\Sigma^3)$ term in the superpotential for
simplicity, since the presence of this term does not change the
conclusions in this paper.

The classical vacua satisfy the following equations:
\begin{equation}
  \label{4}
  \begin{array}{l}
    \hat{m}^A{}_{\! B} \, Q_\alpha^B = \bar{Q}_B^\alpha \,
    \hat{m}^B{}_{\! A} = 0, \\
    \noalign{\vskip 1ex}
    \displaystyle
    m_\Sigma \, \sigmaab = - \lambda \left\{ Q_\alpha^A
    \bar{Q}_B^\alpha - \frac{1}{5} \, \delta^A{}_{\! B} \, Tr(Q \bar{Q})
  \right\},
\end{array}
\end{equation}
where
\begin{equation}
  \label{5}
  \hat{m}^A{}_{\! B} = m \, \delta^A{}_{\! B} + \lambda \sigmaab.
\end{equation}
Together with the $D$-term flatness condition for the gauge group
$\sugut \times \su{3}{H} \times \uone{H}$,
we find four distinct vacua.
Since three of them are not interesting phenomenologically,
we restrict our discussion to one vacuum given by (up to
gauge and global rotations)
\begin{equation}
  \label{7}
  \begin{array}{l}
    \quark = \left(
    \begin{array}{ccc}
      0 & 0 & 0 \\
      0 & 0 & 0 \\
      v & 0 & 0 \\
      0 & v & 0 \\
      0 & 0 & v
    \end{array}
  \right),\
  \antiquark = \left(
  \begin{array}{ccccc}
    0 & 0 & v & 0 & 0 \\
    0 & 0 & 0 & v & 0 \\
    0 & 0 & 0 & 0 & v
  \end{array}
\right), \\
\noalign{\vskip 2ex}
\displaystyle \sigmaab = \frac{m}{\lambda}
\left(
  \begin{array}{ccccc}
    {3 \over 2} & & & & \\
    & {3 \over 2} & & & \\
    & & -1 & & \\
    & & & -1 & \\
    & & & & -1
  \end{array}
\right) ; \quad  v = \sqrt{\frac{5}{2} \frac{m \, m_\Sigma}{\lambda^2}}.
\end{array}
\end{equation}

The vacuum-expectation values in Eq.(\ref{7}) break the original gauge
group down to the standard gauge group:
\begin{equation}
  \label{8}
  \sugut \times \su{3}{H} \times \uone{H} \rightarrow \su{3}{C} \times
  \su{2}{L} \times \uone{Y}.
\end{equation}
Here, $\uone{Y}$ is a linear combination of $\uone{H}$
and a $U(1)$ subgroup of $\sugut$. The GUT unification of three gauge
coupling constants of $\su{3}{C} \times \su{2}{L} \times \uone{Y}$ is
achieved in the limit $g_{1H} \rightarrow \infty$, where $g_{1H}$ is
the gauge coupling constant of the hypercolor $\uone{H}$ \cite{8}.
If one requires the GUT unification by $ 2\%$ accuracy, one gets a
constraint $\alpha_{1 H} \geq 0.06$ for $\alpha_5 \simeq 1/25$ at the
GUT scale (see Ref.\cite{8} for the normalization of $\alpha_{1 H}$) \cite{8A}.

It is amusing that there is no flat direction and hence no massless
state, which results from the fact that
the superpotential (\ref{3}) breaks explicitly the flavor
symmetry (\ref{1}) down to $\sugut \times \uone{H}$. Nambu-Goldstone
multiplets transforming as $(\rep{3}^\ast, \rep{2})$ and $(\rep{3},
\rep{2})$ under $\su{3}{C} \times \su{2}{L}$ are absorbed in the
$\sugut$ gauge multiplets to form massive vector multiplets.

Let us now turn to the $N_f = 6$ case with an additional pair of quark
$\quarksix$ and antiquark $\antiquarksix$, whose mass term is
forbidden by imposing an axial $\uone{A}$ symmetry
\begin{equation}
  \label{9}
  \quarksix \rightarrow e^{i\xi} \quarksix \; \; , \; \;
  \antiquarksix \rightarrow e^{i\xi} \antiquarksix .
\end{equation}
These massless quarks have flat directions since they do not have a
superpotential with $\sigmaab$ and hence there are infinitely
degenerate vacua. Around the vacuum in Eq.(\ref{7}), the flat
directions are given by
\begin{equation}
  \label{10}
  \quarksix = (0, 0, w), \quad
  \antiquarksix =
  \left(
  \begin{array}{c}
    0 \\
    0 \\
    w e^{i\delta}
  \end{array}
\right). \\
\end{equation}
When $w = 0$, the standard gauge group
$\su{3}{C} \times \su{2}{L} \times \uone{Y}$ remains unbroken. We assume
that SUSY-breaking effects choose it as a true vacuum,
avoiding $w \neq 0$ where the color $\su{3}{C}$ would be broken.

It is remarkable that there is a pair of
massless color-triplets $\quarksix$ and $\antiquarksix$
in the above vacuum at $w = 0$. Notice that
the unbroken color $\su{3}{C}$ is a diagonal subgroup of
the hypercolor $\su{3}{H}$ and an $SU(3)$ subgroup of $\sugut$
and the original $\su{3}{H}$ quarks $\quarksix$ and $\antiquarksix$
transform as $\rep{3}$ and $\rep{3}^\ast$ under the color
$\su{3}{C}$.
The presence of massless color-triplets $\quarksix$ and $\antiquarksix$ is a
crucial point for obtaining a pair of light Higgs doublets as seen in
section 4.

Before proceeding to the quantum analysis of the vacuum chosen above,
two remarks are in order here:

$i)$
The classical moduli space of vacua has a hypercolor-gauge invariant
description in terms of the observable ``meson'' $M$
and ``baryons'' $B$ and $\bar{B}$:
\begin{eqnarray}
  \label{12}
  & & \Meson = \quark \bar{Q}^\alpha_B , \nonumber \\
  & & B^{[ABC]} = \frac{1}{3!} \epsilon^{\alpha \beta \gamma} Q^A_\alpha
  Q^B_\beta Q^C_\gamma, \\
  & & \bar{B}_{[ABC]} = \frac{1}{3!} \epsilon_{\alpha \beta \gamma}
  \bar{Q}_A^\alpha \bar{Q}_B^\beta \bar{Q}_C^\gamma \nonumber .
\end{eqnarray}

The vacua corresponding to Eqs.(\ref{7}) and (\ref{10}) are given by
\begin{equation}
  \label{13}
  \begin{array}{l}
    \displaystyle \Meson = \left(
    \begin{array}{cccccc}
      0 & & & & & \\
      & 0 &  & & & \\
      & & v^2 & & & \\
      & & & v^2 & & \\
      & & & & v^2 & vw e^{i \delta} \\
      & & & & vw  & w^2 e^{i \delta}
    \end{array}
  \right) , \\
  \noalign{\vskip 2ex}
  B^{[3,4,5]} = \bar{B}_{[3,4,5]} = v^3 , \\
  \noalign{\vskip 1ex}
    B^{[3,4,6]} = v^2 w, \quad
    \bar{B}_{[3,4,6]} = v^2 w e^{i \delta}
\end{array}
\end{equation}
with all the other components of $B$ and $\bar{B}$ vanishing.

It is clear
that the flavor gauge group $\sugut \times \uone{H}$ is spontaneously
broken down to $\su{3}{C} \times \su{2}{L} \times \uone{Y}$ in the
$w = 0$ vacuum. Then we have a
pair of massless composite states
\begin{equation}
  \label{14}
  M^6{}_{\! a} = \quarksix \bar{Q}^\alpha_a \; \; , \; \; M^a{}_{\!
    6} = Q^a_\alpha \bar{Q}^\alpha_6 \quad (a = 3, \cdots, 5),
\end{equation}
which are $\rep{3}$ and $\rep{3}^\ast$ of the color $\su{3}{C}$,
respectively.

$ii)$
We note that there is another interesting vacuum in the present model:
\begin{equation}
  \label{11}
  \begin{array}{l}
    \quark = \left(
    \begin{array}{ccc}
      0 & 0 & 0 \\
      0 & 0 & 0 \\
      0 & 0 & 0 \\
      0 & v & 0 \\
      0 & 0 & v \\
      0 & 0 & 0
    \end{array}
  \right),\
  \antiquark = \left(
  \begin{array}{cccccc}
    0 & 0 & 0 & 0 & 0 & 0 \\
    0 & 0 & 0 & v & 0 & 0 \\
    0 & 0 & 0 & 0 & v & 0
  \end{array}
\right), \\
\noalign{\vskip 2ex}
\displaystyle \sigmaab = \frac{m}{\lambda}
\left(
  \begin{array}{ccccc}
    {2 \over 3} & & & & \\
    & {2 \over 3} & & & \\
    & & {2 \over 3}  & & \\
    & & & -1 & \\
    & & & & -1
  \end{array}
\right) \, ; \quad v = \sqrt{\frac{5}{3} \frac{m \, m_\Sigma}{\lambda^2}}.
\end{array}
\end{equation}

In this vacuum $\uone{Y}$ is a diagonal subgroup of $U(1)$'s in
$\sugut$ and $\su{3}{H}$ and hence the additional $\uone{H}$ is not
necessary.
However, unwanted massless states $Q^6_1$ and $\bar{Q}^1_6$ exist
in addition to a pair of $\su{2}{L}$ doublets $\quarksix$ and
$\antiquarksix$ $(\alpha = 2 , 3)$ which may be identified with the
Higgs multiplets in the standard model.
These massless multiplets $Q^6_1$ and $\bar{Q}^1_6$ have $\uone{Y}$
charges $-1$ and $+1$, respectively, and give an additional
contribution to the renormalization-group equations for gauge coupling
constants. This destroys the success of gauge coupling unification in
the SUSY-GUT. Thus we do not investigate this vacuum in this paper
\cite{8a}.

\section{Quantum vacua and an effective superpotential}

Let us analyze quantum effects on the classical vacua given in
Eqs.(\ref{7}) and (\ref{10}).
We see that the
effective mass matrix for quarks has four zero eigenvalues
in our vacua:
\begin{equation}
  \label{15}
  \hat{m}' =
  \left(
    \begin{array}{cccccc}
      & & & & & \\
      & \mbox{\LARGE $\hat{m}$} & & & & \\
      & & & & & \\
      & & & & & \\
      & & & & & 0
    \end{array}
  \right)
  =
  \left(
    \begin{array}{cccccc}
      \frac{5}{2}m & & & & \\
      & \frac{5}{2}m & & & \\
      & & 0 & & & \\
      & & & 0 & & \\
      & & & & 0 & \\
      & & & & & 0
    \end{array}
  \right).
\end{equation}
Therefore, our model becomes a SUSY QCD-like theory with the
effective $N_f = 4$ at low energies.

Expanding the $\Sigma$ fields around the values $\langle \Sigma
\rangle$ given in Eq.(\ref{7}), we write the tree-level superpotential
as
\begin{equation}
  \label{a1}
  \begin{array}{ll}
    W = & \bar{Q}^\alpha_i \left( \displaystyle \frac{5}{2} \, m \,
    \delta^i{}_{\! j} + \lambda \sigma^i{}_{\! j} \right) Q^j_\alpha
    + \lambda \, \bar{Q}^\alpha_a \left(
    \sigma^a{}_{\! b} \right) Q^b_\alpha
    + \lambda \, \bar{Q}^\alpha_a \left(
    \sigma^a{}_{\! i} \right) Q^i_\alpha + \lambda \,
    \bar{Q}^\alpha_i \left(
    \sigma^i{}_{\! a} \right) Q^a_\alpha \\
    \noalign{\vskip 0.5ex}
    & \displaystyle + \frac{\lambda}{\sqrt{30}} \, \sigma_0 \left( 3
    \bar{Q}^\alpha_i Q^i_\alpha - 2 \bar{Q}^\alpha_a
    Q^a_\alpha \right)
    \displaystyle + \frac{1}{2} \, m_\Sigma \, \left\{ Tr(\sigma^2)
    + \sigma_0^2 \right\} +  \frac{\sqrt{30}}{2} \, \frac{m \,
      m_\Sigma}{\lambda} \, \sigma_0 \\
    \noalign{\vskip 1ex}
    & \hspace{8em} (i, j = 1, 2 ; \; \;  a, b = 3, \cdots, 5) \; ,
  \end{array}
\end{equation}
where $\sigma$ and $\sigma_0$ are defined by
\begin{equation}
  \label{a2}
  \begin{array}{c}
    \displaystyle \Sigma^A{}_{\! B} = \langle \Sigma^A{}_{\! B}
    \rangle + \sigma^A{}_{\! B} + \frac{1}{\sqrt{30}}
    \left(
      \begin{array}{ccccc}
        3 & & & & \\
        & 3 & & & \\
        & & -2 & & \\
        & & & -2 & \\
        & & & & -2
      \end{array}
    \right)
    \sigma_0  \, , \\
    \noalign{\vskip 1ex}
    \sigma^A{}_{\! B} =
    \left(
      \begin{array}{cc}
        \sigma^i{}_{\! j} & \sigma^i{}_{\! b} \\
        \sigma^a{}_{\! j} & \sigma^a{}_{\! b} \\
      \end{array}
    \right)\, ; \quad Tr \, \sigma^i{}_{\! j} = Tr \, \sigma^a{}_{\!
      b} = 0 \, \\
    \noalign{\vskip 1ex}
    \; \; \; \; \; \; \; \; (A, B = 1, \cdots, 5).
  \end{array}
\end{equation}
Notice that there is a tadpole term for $\sigma_0$ in
Eq.(\ref{a1}), which is canceled out by the $\bar{Q}^\alpha_a \,
Q^a_\alpha$ condensation.

We now think of
our model as the QCD-like theory with two massive ($Q^i_\alpha$
and $\bar{Q}^\alpha_i$) and four massless ($Q^a_\alpha \; {\rm and} \;
\bar{Q}^\alpha_a \; ; \, a = 3, \cdots, 6$)
quarks interacting with the $\sigma$ fields.
Then we can integrate out the massive quarks
and irrelevant $\sigma$ fields \cite{8b}
to get the low-energy effective theory described by four massless quarks
with the superpotential
\begin{equation}
  \label{a4}
  \begin{array}{ll}
    W_{low} = & \displaystyle \lambda \, \bar{Q}^\alpha_a
    \left( \sigma^a{}_{\! b} - \frac{2}{\sqrt{30}} \, \sigma_0
      \, \delta^a{}_{\! b}
    \right) Q^b_\alpha \\
    \noalign{\vskip 1ex}
    & \displaystyle + \frac{1}{2} \, m_\Sigma \left\{ Tr(\sigma^2)
    + \sigma_0{}^2 \right\}
    + \frac{\sqrt{30}}{2} \, \frac{m \, m_\Sigma}{\lambda} \sigma_0 \\
    \noalign{\vskip 1ex}
    & \hspace{8em} (a, b = 3, \cdots, 5) \; .
  \end{array}
\end{equation}
Here we have eliminated the Nambu-Goldstone modes stemming from
breakdown of the flavor symmetry, since they will be absorbed in the
gauge multiplets of $\sugut$ to form massive vector multiplets.

We next proceed to find out the effective superpotential described by
the composite meson $M$ and baryons $B$ and $\bar{B}$ interacting with
the $\sigma$ fields. It is highly nontrivial to obtain the
superpotentials dynamically generated by strong interactions. However,
it has become clear recently that the effective superpotentials can be
exactly determined for certain classes of SUSY nonabelian gauge
theories \cite{10,12}.
In particular, for the QCD-like theory with $N_f = N_c +1$
flavors of quarks where $N_c$ is the number of colors, the quantum
moduli space of vacua is the same as the classical one and the
effective superpotential at low energies is uniquely determined
\cite{12}.

According to Ref.\cite{10,12}, we
obtain the effective superpotential for our model from Eq.(\ref{a4})
as
\begin{equation}
  \label{16}
  \begin{array}{lll}
  W_{eff} & = & \displaystyle W_{dyn} + \lambda \, Tr(M
  \tilde{\sigma}) + \frac{1}{2} \, m_\Sigma
  \left\{ Tr(\sigma^{\, 2}) + \sigma_0{}^2
  \right\}
  + \frac{\sqrt{30}}{2} \, \frac{m \, m_\Sigma}{\lambda} \, \sigma_0;
  \\
  \noalign{\vskip 1ex}
  W_{dyn} & = & \Lambda^{-5} ( \baryon \meson \antibaryon - \det M ),
  \end{array}
\end{equation}
where $\meson$ , $\baryon$ and $\bar{B}^a$ are composite meson and baryon
states in the $N_f = 4$ case:
\begin{eqnarray}
  \label{16a}
  \begin{array}{lll}
    & & \meson \sim Q^a_\alpha \bar{Q}^\alpha_b , \\
    \noalign{\vskip 1ex}
    & & \displaystyle \baryon \sim \left( \frac{1}{3!} \right)^2
    \epsilon^{\alpha \beta \gamma} \epsilon_{a b c d} Q^b_\alpha
    Q^c_\beta Q^d_\gamma , \\
    \noalign{\vskip 1ex}
    & & \displaystyle \bar{B}^a \sim \left( \frac{1}{3!} \right)^2
    \epsilon_{\alpha \beta \gamma} \epsilon^{a b c d} \bar{Q}_b^\alpha
    \bar{Q}_c^\beta \bar{Q}_d^\gamma \\
    \noalign{\vskip 1ex}
    & & \hspace{7em} (a, b = 3, \cdots , 6) \, .
  \end{array}
\end{eqnarray}
$\Lambda$ denotes the dynamical scale of $\su{3}{H}$, and
$\tilde{\sigma}$ is defined by
\begin{equation}
  \label{17}
  \tilde{\sigma} = \left(
  \begin{array}{cc}
    \displaystyle \sigma^a{}_{\! b} & \\
    & 0
  \end{array} \right)
  - \frac{2}{\sqrt{30}} \left(
  \begin{array}{cccc}
    1 & & & \\
    & 1 & & \\
    & & 1 & \\
    & & & 0
  \end{array} \right) \sigma_0 \, .
\end{equation}

{}From Eq.(\ref{16}) we find that the quantum vacua satisfy
\begin{equation}
  \label{18}
  \begin{array}{l}
    \meson \antibaryon = \baryon \meson = 0,  \\
    \noalign{\vskip 1ex}
    \displaystyle
    -\frac{\partial \det M}{\partial \meson} + \baryon \antibaryon +
    \Lambda^5 \lambda \tilde{\sigma}^b{}_{\! a} = 0 \; , \\
    \noalign{\vskip 1ex}
  \end{array}
\end{equation}
for $a, b = 3, \cdots, 6$;
\begin{equation}
  \label{18A}
  \begin{array}{l}
    \noalign{\vskip 1ex}
    \displaystyle
    m_\Sigma \, \sigma^a{}_{\! b} + \lambda \left \{
    \meson - \delta^a{}_{\! b} \, Tr (\meson) \right \} = 0, \\
    \noalign{\vskip 1ex}
    \displaystyle
    m_\Sigma \, \sigma_0 + \frac{\sqrt{30}}{2} \, \frac{m \,
      m_\Sigma}{\lambda} - \frac{2}{\sqrt{30}} \, \lambda \, Tr
    (\meson) = 0
    \; ,
  \end{array}
\end{equation}
for $a, b = 3, \cdots, 5$.
Using the $D$-term flatness condition for the flavor gauge group
$\sugut \times \uone{H}$ together with Eqs.(\ref{18}) and (\ref{18A}),
we obtain the quantum vacua which are exactly the same
as those given in Eq.(\ref{13}) for the classical theory. We take, as
before, the $\su{3}{C} \times \su{2}{L} \times \uone{Y}$ invariant
vacuum at $w = 0$.

In this vacuum, the mass matrix for color-triplet states $M$, $B$ and
$\bar{B}$ is given by
\begin{equation}
  \label{18a}
  \begin{array}{l}
    \begin{array}{cc}
      \vspace{3ex}
      ( M^6{}_a & \baryon ) \\
      \vphantom{0}
    \end{array}

    \left( \begin{array}{cc}
      \displaystyle \frac{v^4}{\Lambda^5} & \displaystyle -
      \frac{v^3}{\Lambda^5} \\
      \noalign{\vskip 0.5ex}
      \displaystyle - \frac{v^3}{\Lambda^5} & \displaystyle
      \frac{v^2}{\Lambda^5}
    \end{array} \right)

    \left( \begin{array}{c}
      \vspace{3.5ex}
      M^a{}_{\! 6} \\
      \bar{B}^a
    \end{array} \right) \; . \\
  \end{array}
\end{equation}
Notice that $M$ and $B \, (\bar{B})$ have canonical dimensions two
and three, respectively.
A pair of massless color-triplet states is now a mixture of $M, B$
and $\bar{B}$ with its orthogonal states
having GUT scale masses. We will see later that the structure of the
mass matrix for these color-triplet states is crucial for suppression
of the dangerous dimension-five operators for nucleon decays.

The Nambu-Goldstone modes in $\meson \; (a, b = 3, \cdots, 5)$
get tree-level
masses through the interactions (the $\lambda$ coupling term)
with $\sigma$ and $\sigma_0$ given in
Eq.(\ref{16}). Thus there are no other
massless composite states besides the $\su{3}{C}$ triplets
observed in the mass matrix (\ref{18a}).

\section{Missing partner mechanism for light Higgs doublets}

We introduce a standard Higgs multiplets, $H_A$ and $\bar{H}^A$ (with
$A = 1, \cdots, 5$), in the fundamental representations $\rep{5}$ and
$\rep{5}^\ast$ of $\sugut$. The mass term for them is forbidden by the
axial $\uone{A}$ symmetry given in Eq.(\ref{9}). Assuming their
transformation property under $\uone{A}$ as
\begin{equation}
  \label{21}
  \higgs \rightarrow e^{- i \xi} \higgs \; , \; \antihiggs \rightarrow
  e^{-i \xi} \antihiggs ,
\end{equation}
we have a superpotential
\begin{equation}
  \label{22}
  W_H = h \higgs \quark \antiquarksix + h' \antihiggs \antiquark
  \quarksix .
\end{equation}

These Yukawa interactions induce the following mass terms at low
energies:
\begin{equation}
  \label{23}
  W_{mass} = h H_a M^a{}_{\! 6} + h' \bar{H}^a M^6{}_{\! a}
\end{equation}
for $a = 3, \cdots, 5$. Thus the color-triplet components of $\higgs$ and
$\antihiggs$ denoted by $H_a$ and $\bar{H}^a$ $(a = 3, \cdots, 5)$ become
massive together with the massless composites in $M^6{}_{\! a}$,
$M^a{}_{\! 6}$,  $\baryon$
and $\bar{B}^a$.
On the other hand, the doublet components of $\higgs$
and $\antihiggs$ denoted by $H_i$ and $\bar{H}^i$ $(i = 1, 2)$
remain massless.

This missing partner mechanism is easily understood in terms of the
Higgs phase, where $Q^a_\alpha$ and $\bar{Q}^\alpha_a$ have
vacuum-expectation values for $a = 3, \cdots, 5$ as shown in
Eq.(\ref{7}). Then, it is clear from Eq.(\ref{22}) that the
color-triplets $H_a$ and $\bar{H}^a$ get masses together with
$\antiquarksix$ and $\quarksix$, respectively. On the other hand, the
$\su{2}{L}$-doublet Higgses $H_i$ and $\bar{H}^i$ have no partners to
form masses and the $H_i \bar{H}^i$ mass term itself is forbidden by
the axial $\uone{A}$ in Eq.(\ref{21}).
However, we need a more careful
analysis at the quantum level, since the axial $\uone{A}$ is broken by
the hypercolor $\su{3}{H}$ instantons.

We now prove that the Higgs doublets $H_i$ and $\bar{H}^i$ are exactly
massless even at the quantum level in the limit of SUSY being
exact.
We first integrate out the massive quarks $Q^i_\alpha$ and
$\bar{Q}^\alpha_i$ to get the low-energy
superpotential (neglecting the irrelevant $\sigma$ fields)
\begin{equation}
  \label{29}
  W'_{low} = W_{low} + h H_a Q^a_\alpha \antiquarksix + h' \bar{H}^a
  \bar{Q}^\alpha_a \quarksix + \frac{2 h h'}{5m} H_i \bar{H}^i \quarksix
  \antiquarksix \, ,
\end{equation}
where $W_{low}$ is given in Eq.(\ref{a4}).
Using the methods proposed in Ref.\cite{10,12}, we obtain the
effective superpotential
\begin{equation}
  \label{30}
  W'_{eff} = W_{eff} + h H_a M^a{}_{\! 6} + h' \bar{H}^a M^6{}_{\! a}
  + \displaystyle \frac{2 h h'}{5m} H_i \bar{H}^i M^6{}_{\! 6} \; ,
\end{equation}
where $W_{eff}$ is given in Eq.(\ref{16}).
Since $M^6{}_{\! 6}$ vanishes in the present vacuum, no mass term for
the Higgs doublets $H_i$ and $\bar{H}^i$ is generated.

This important conclusion is also understood if one notices that the
hypercolor-anomaly (instanton) effects are independent of any Yukawa
coupling constants $\lambda$, $h$ and $h'$, and hence they are present
only in the dynamical part $W_{dyn}$ in $W'_{eff}$. Therefore, the
Yukawa-coupling dependent parts of $W'_{eff}$ must be
$\uone{A}$-invariant, which shows that the $H_i \bar{H}^i$ terms are
always accompanied by $M^6{}_{\! 6}$ as seen in Eq.(\ref{30}).
On the other hand, the dynamical part $W_{dyn}$ does not have the $H_i
\bar{H}^i$ term. This is easily proved by means of a higher symmetry in
the limit of $h = h' = 0$ ({\it e.g.} $H_i \rightarrow e^{i \, \alpha}
H_i$, $\bar{H}^i \rightarrow e^{i \, \beta} \bar{H}^i$ and all the
other fields intact).

Note that the sixth quarks will become massive when the Higgs doublets
$H_i$ and $\bar{H}^i$ acquire the vacuum expectation values at the
electroweak scale, which is consistent with the choice $w = 0$ in
Eq.(\ref{13}).

\section{Discussions}

We have shown in this paper that our QCD-like theory with six quark
flavors generates one pair of massless Higgs doublets
naturally. These Higgs doublets survive all the quantum
corrections and remain in the massless spectrum. The masslessness of
the original Higgs multiplets, $\higgs$ and $\antihiggs \; (A = 1, \cdots,
5)$, is understood by the axial $\uone{A}$ symmetry. Although this
axial $\uone{A}$ is broken by the hypercolor
instantons, the presence of massless Higgs doublets has
been proved by means of
the nonrenormalization theorem in SUSY theories \cite{10,12}.

It is remarkable that nucleon decays due to the dangerous dimension-five
operators are suppressed in our model. These operators are
induced by exchanges of $\su{3}{C}$ triplet Higgs multiplets $H_a$ and
$\bar{H}^a$ \cite{20}. Owing to (\ref{18a}) and (\ref{23}),
the mass matrix for the color-triplet sector is given by
\begin{equation}
  \label{26}
    \begin{array}{ccc}
      \vphantom{0} \\
      ( H_a & M^6{}_a & \baryon ) \\
      \vphantom{0}
    \end{array}
    \hat{m}_C
    \left( \begin{array}{c}
      \bar{H}^a \\ M^a{}_{\! 6} \\ \bar{B}^a
    \end{array} \right),
\end{equation}
where
\begin{equation}
  \label{26a}
    \hat{m}_C =
    \left( \begin{array}{ccc}
      \displaystyle 0 & h & 0 \\
      \noalign{\vskip 0.5ex}
      h' & \displaystyle \frac{v^4}{\Lambda^5} & \displaystyle -
      \frac{v^3}{\Lambda^5} \\
      \noalign{\vskip 0.5ex}
      0 & \displaystyle - \frac{v^3}{\Lambda^5} & \displaystyle
      \frac{v^2}{\Lambda^5}
    \end{array} \right).
\end{equation}
The dimension-five operators are proportional to $\left( \hat{m}_C^{-1}
\right)_{11}$, which is none other than zero.
This conclusion can also be understood in terms of the Higgs phase:
the mass for $\quarksix$ and $\antiquarksix$ is not generated even in
the presence of instanton effects and hence there is no transition
matrix between $H_a$ and $\bar{H}^a$. We note that this is consistent
with the result by Affleck, Dine and Seiberg \cite{11}.

The GUT unification of three gauge coupling constants of
$\su{3}{C} \times \su{2}{L} \times \uone{Y}$ is achieved in the strong
coupling limit of $\uone{H}$. The necessity of the strong $\uone{H}$ is
a possible drawback in the present model. However, it is very hard to
distinguish phenomenologically the standard SUSY-GUT and our model,
since the $\su{3}{C}$ gauge coupling constant has experimental errors
of several percent (see Ref.\cite{8}). Moreover, some threshold
corrections from GUT-scale particles also give a few percent
ambiguity to the renormalization-group equations \cite{18}.

{}From a theoretical point of view, it may be a problem that the
electromagnetic charge quantization is not an automatic consequence in
our model. However, this can be solved by assuming the hypercolor
$\su{3}{H} \times \uone{H}$ to be embedded in, for example, $\su{4}{H}$
at some higher scale. This extension of the hypercolor group also
solves the problem that the gauge coupling constant of $\uone{H}$
blows up below the Planck scale. This possibility together with a
phenomenological analysis will be considered in a future communication
\cite{19}.

In the present paper we have assumed the global SUSY theory.
In the framework of supergravity, the flat directions
($w$ and $w e^{i\delta}$) in Eq.(\ref{10}) may be lifted substantially
when they reach the Planck scale. However, effects from such vacuum shifts
are negligibly small near the origin ($w = 0$) of the flat directions.

In our model the global $\uone{A}$ plays a central role for having
a pair of massless Higgs doublets. Such a global symmetry
may be, in general, broken by topology-changing wormhole effects \cite{21}.
If it is the case, the pair of Higgs doublets is no longer massless.
The magnitudes of these induced operators are, however, not known and
hence we assume that these effects are sufficiently suppressed.

Finally we make a comment on the basic structure of our model. We have
assumed a mass term for the first five quarks as seen in
Eq.(\ref{3}). However, we can obtain the same result by introducing a
coupling with a singlet field $\phi$ instead of the mass term.
In such a model the
vacuum-expectation value of $\phi$ plays a role of the mass $m$. This
model seems very intriguing,
since it may be regarded as a massless QCD-like theory
with nonrenormalizable interactions
if one integrates out the $\phi$ and $\Sigma$ multiplets.


\begin{thebibliography}{99}
\bibitem{8A} The color $\su{3}{C}$ is also a linear combination of the
  hypercolor $\su{3}{H}$ and an $SU(3)$ subgroup of $\sugut$. Thus a
  similar consideration applies for $\su{3}{H}$, whose coupling is
  strong enough under our setting of the confining $\su{3}{H}$ at the GUT
  scale.
\bibitem{8a} If one introduces a pair of $\rep{10}$ and
  $\rep{10}^\ast$ of $\sugut$ and assumes a superpotential $W =
  \frac{\lambda'}{M} \epsilon^{\alpha \beta \gamma} \quark Q^B_\beta
  Q^6_\gamma \, \rep{10}_{AB} + \frac{\lambda''}{M} \epsilon_{\alpha
    \beta \gamma} \antiquark \bar{Q}^\beta_B \bar{Q}^\gamma_6 \,
  \rep{10}^{\ast AB}$, the unwanted singlets $Q^6_1$ and $\bar{Q}^1_6$
  may have large masses in the vacuum Eq.(\ref{11}). However, we do
  not pursuit such a possibility here.
\bibitem{8b} The equations of motion for massive quarks lead to
  $Q^i_\alpha = \bar{Q}^\alpha_i = 0 \; \; \; \; (i = 1, 2)$.
  From the Konishi-anomaly relation \cite{9}, we see that
  the massive quarks have vanishing condensations, {\it i.e.}
  $\bar{Q}^\alpha_i Q^j_\alpha = 0$ even at the quantum level,
  which are consistent with the results implied by
  the equations of motion. We note that they are also obtained
  by means of the dual theory \cite{16} to our model.
\end{thebibliography}
\end{document}